\documentclass{elsart}
\usepackage{ifpdf}
\usepackage{graphicx,amssymb,lineno}
\usepackage{amssymb,lineno,amsfonts}
\usepackage{graphicx}   % Include figure files
\usepackage{dcolumn}    % Align table columns on decimal point
\usepackage{bm}         % bold math
\usepackage{bbm}
\usepackage{mathrsfs}
\usepackage{upgreek}
\usepackage{mathtools}

\begin{document}
\begin{frontmatter}
\title{Characterization of RF Inhomogeneity via NOON states}

\author{Abhishek Shukla}
and
\author{T. S. Mahesh}$^*$
\ead{*mahesh.ts@iiserpune.ac.in}
\address{Department of Physics and NMR Research Center, \\
				 Indian Institute of Science Education and Research,
				 Pune 411 008.}
\date{\today}

\begin{abstract}
{We describe a method for quantitative characterization of radio-frequency (RF)
inhomogeneity using NOON states.  NOON states are special multiple-quantum
coherences which can be easily prepared in star-topology spin-systems.  
In this method we exploit the high sensitivity of the NOON states for
z-rotations.
As a result, Torrey oscillations with NOON states decay much faster than that of single
quantum coherences. Therefore the present method requires shorter pulse durations,
and enables the study of inhomogeneity at higher RF powers.  
To model such a inhomogeneity profile, we propose a two-parameter asymmetric Lorentzian
function.  The experiments are carried out in $^1$H channels of two different 
high-resolution NMR probes and the results are compared.
Finally, we also extend the NOON state method to characterize the inhomogeneity correlations
between two RF channels.  We obtain the correlation profile between $^1$H and $^{31}$P channels 
of an NMR probe by constructing a five-parameter asymmetric-3D Lorentzian model.
}      
\end{abstract}

\end{frontmatter}

\section{Introduction} 
The strength of NMR over other spectroscopy techniques is in the excellent control over
quantum dynamics \cite{levittbook}.  
Coherent control of nuclear spins is achieved by a calibrated set of radio frequency pulses.  
Due to its low sensitivity, NMR needs an ensemble of spatially distributed replicas 
of a spin system. NMR signal is the ensemble average of signals arising from each of 
these replicas.  Usually, it is assumed that each member of the ensemble undergoes
the same unitary dynamics.  In practice, however, RF amplitude has a spatial distribution
over the sample volume.  This is usually referred to as RF inhomogeneity (RFI)  \cite{cavanagh}.

Estimation of RFI in a given NMR probe is important not only to characterize the 
quality of the probe, but also for experiments requiring precise control of 
spin-dynamics. 
RFI characterization has helped designing robust and high-fidelity 
quantum gates for quantum information studies \cite{coryrfijcp}.  
In MRI, RFI characterization can help to understand certain image distortions and to
correct them \cite{mri1,mri2}.

\begin{figure}[b]
\begin{center}
\includegraphics[width=9cm]{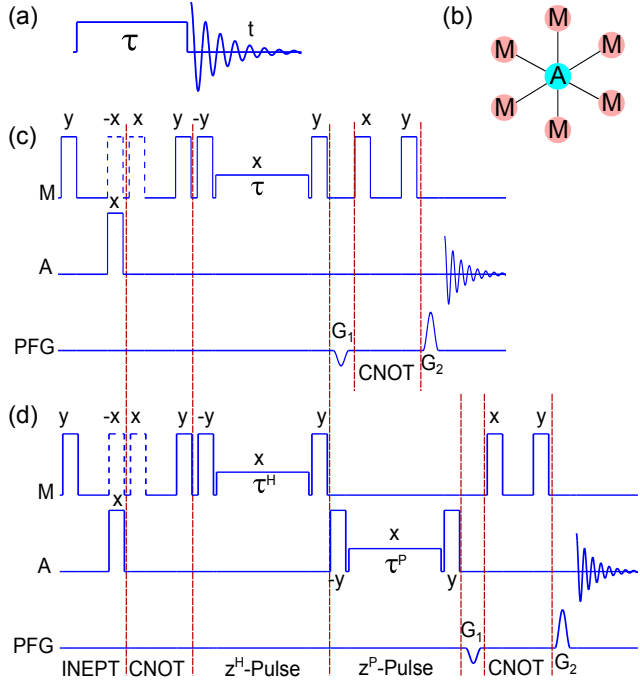}
\caption{(a) Pulse sequence for single-quantum Torrey oscillation,
(b) star-topology spin-systems, (c) pulse sequence for NOON Torrey oscillation, and
(d) pulse sequence for the measurement of correlation between two RF channels.
In (c) and (d), except the variable duration ($\tau$) pulses, all other pulses are 90$^\circ$
pulses.  The operators corresponding to the dashed pulses cancel each other and can be omitted.   
}
\label{torrey}
\end{center}
\end{figure}

The standard method for the measurement of RFI is via Torrey oscillation \cite{torrey}.  
A single, on-resonant, low-power RF pulse of variable duration is applied on a 
spin-1/2 system with suitably long relaxation time constants.  The pulse diagram is shown
in Fig. \ref{torrey}(a).
Ideally, the magnetization
should nutate, without decay, about the direction of RF.  But because of RF inhomogeneity,
the magnetizations at different parts of the sample nutate with slightly different frequencies,
ultimately resulting in the decay of the overall magnetization.  This decaying spiral in vertical
plane perpendicular to the direction of RF is known as Torrey oscillation.  Since NMR
signal is always proportional to the transverse component of the magnetization, the amplitude of
the signal collected at different points of the Torrey oscillation also displays oscillatory decay.  
The Fourier transform of this decay can be used to characterize RFI.  While this method seems
to be the simplest way to characterize RFI, the main disadvantage is the necessity of long
RF pulses to allow complete decay of Torrey oscillation.  Fourier transform of incomplete 
decays will lead to artifacts.  In a good NMR probe, Torrey oscillation at the maximum allowed RF power
may last longer than several milli-seconds and exceeds the duty-cycle limit. 
This makes the measurement of RFI at high powers prohibitive.  

Here we describe a method to speed-up the decay of Torrey oscillations using NOON states.  
NOON states are the highest order multiple-quantum coherences in a given spin-system.  
%The fast decay of Torrey oscillations enables one to
%study RFI even at higher RF powers.
Previously, NOON states have been used in NMR for ultra-low magnetic field sensing \cite{jonesscience} and for 
diffusion studies \cite{abhishekdiff}.  

In the following section we describe the theory of Torrey oscillation, preparation of NOON states, 
the measurement of RFI using NOON states, and characterization of RFI correlations between two
channels of a high resolution NMR probe.  The experimental details are described in section 3,
and finally we conclude in section 4.

\section{Theory}
\subsection{Radio frequency inhomogeneity}
For a given nominal RF amplitude $\nu_0$, let us assume RFI distribution $\{\nu, p(\nu) \}$, 
where $p(\nu)$ is the probability, in the sample volume, of the RF amplitude $\nu$.  The mean
RF amplitude is
%\begin{eqnarray}
$\overline{\nu} = \sum_\nu p(\nu) \nu.$
%\end{eqnarray}
Here we have defined the nominal RF amplitude $\nu_0$ as the one with the highest probability.
In an asymmetric RFI distribution, $\nu_0$ may be different from $\overline{\nu}$.  

Consider the standard experiment for estimating RFI via Torrey oscillation as shown in Fig. \ref{torrey}(a).
During an on-resonant pulse, the distribution of nutation frequencies $\nu_1$ of the single quantum coherence 
is same as the RFI distribution $p(\nu)$.  
%Therefore the nutation frequency has a distribution
%$\{\nu_1, p(\nu_1)\}$.
Therefore real part of the signal collected at the end of a variable-duration ($\tau$) y-pulse is
$s(\tau) e^{-t/T_2}$, where
\begin{eqnarray}
s(\tau) = \sum_{\nu} p(\nu) \sin(2\pi \nu \tau).
\end{eqnarray}
Here $T_2$ is the spin-spin relaxation time constant.  
Similarly, for a $q$-quantum coherence
\begin{eqnarray}
s(\tau) = \sum_{\nu} p(\nu) \sin(2\pi q \nu \tau).
\end{eqnarray}
The Fourier transform along $t$ of $s(\tau) e^{-t/T_2}$  will result
in a Lorentzian of line-width $\pi/T_2$ centered at zero frequency (on-resonant case)\cite{mathbook}.  
The amplitude of this Lorentzian,
is proportional to $s(\tau)$. We record a series of experiments by systematically incrementing $\tau$ and obtain
the Lorentzian peak-heights $s(\tau)$. 
In a typical NMR probe, the real part of the Fourier transform
% The probe RFI can be characterized from 
%the normalized real part of Fourier transform of $A(\tau)$,
\begin{eqnarray}
S(\nu_q) = {\cal F}_r\left[ s(\tau) \right]
\label{snu}
\end{eqnarray}
displays an asymmetric profile, with a maximum at the nominal RF frequency $\nu_0$ \cite{coryrfijcp}.
This is because in most cases, the probability of finding lower-than-nominal amplitudes is
more compared to higher-than-nominal.
This profile of RF amplitudes can be modeled by an asymmetric Lorentzian
\begin{eqnarray}
p(\nu) &=& \frac{\lambda_1^2}{(1 - \frac{\nu}{\nu_0})^2+\lambda_1^2} ~~ \mathrm{if} ~~ \nu <   \nu_0, ~~ \mathrm{and,} \nonumber \\
p(\nu) &=& \frac{\lambda_2^2}{(1 - \frac{\nu}{\nu_0})^2+\lambda_2^2} ~~ \mathrm{if} ~~ \nu \ge \nu_0.
\label{rfimodel}
\end{eqnarray}
The area of $p(\nu)$ is normalized to unity.

Using such a model for RFI it is possible to reconstruct $q$-quantum Torrey oscillations
\begin{eqnarray}
s_m(\tau) = \sum_{\nu = 0}^{\nu = 2 \nu_0} p(\nu) \sin(2\pi q \nu \tau),
\label{sm}
\end{eqnarray}
where $\nu$ is the RF frequency with probability $p(\nu)$.
The positive real part of the Fourier transform of $s_m(\tau)$ gives the model
profile $S_m(\nu_q)$.
A scalar measure of distance between the model and the experimental profiles 
is provided by the norm $\vert \vert S(\nu_q) - S_m(\nu_q) \vert \vert$.  
The RFI parameters $(\lambda_1,\lambda_2)$ can be obtained by minimzing the above norm
using a non-linear fit.

For the hypothetical case of $\lambda_1 = \lambda_2 := \lambda$, one can model the Torrey oscillation as
%\begin{eqnarray}
${\cal F}^{-1}_r\left[S_m(\nu_q)\right] \propto \exp(-\lambda \nu_0 q \tau).$
%$\label{Avsv0}
%\end{eqnarray}
Therefore increase in the nominal RF frequency $\nu_0$ will correspondingly speed-up the decay of
Torrey oscillation.  Another way to speedup the decay of Torrey oscillation is to exploit multiple-quantum
coherences.  In the following we explain how NOON states can be used for efficient characterization of RFI.

\subsection{NOON state preparation}
We label the Zeeman eigenstates of a single spin-1/2 nucleus as 
$\left\{\vert 0 \rangle, \vert 1 \rangle \right\}$.
In an $N$-spin system, the NOON state is the superposition of a state $\vert N, 0 \rangle$
(with $N$ spins in $\vert 0 \rangle$ and $0$ spins in $\vert 1 \rangle$) and the state 
$\vert 0, N \rangle$
(with $0$ spins in $\vert 0 \rangle$ and $N$ spins in $\vert 1 \rangle$),
\begin{eqnarray}
\vert N00N \rangle &=& (\vert N, 0 \rangle + \vert 0, N \rangle)/\sqrt{2} \nonumber \\
                   &=& (\vert 00 \cdots 0 \rangle + \vert 11 \cdots 1 \rangle)/\sqrt{2}.
\end{eqnarray}

\begin{figure}[b]
\begin{center}
\includegraphics[width=11cm]{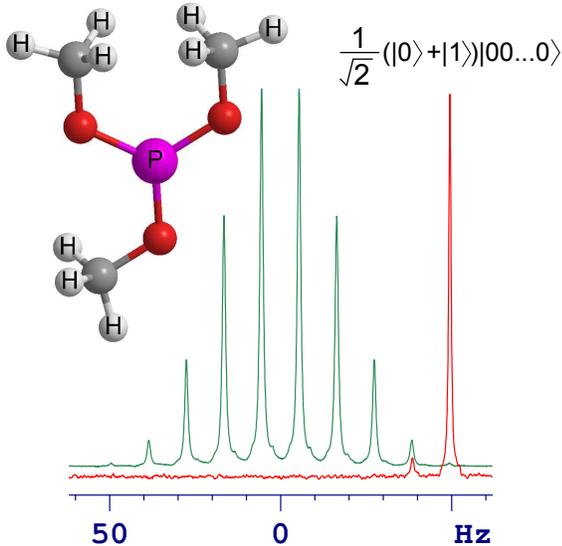}
\vspace*{-2cm}
\caption{Molecular structure of trimethylphosphite and its $^{31}$P spectra
in equilibrium (upper trace) and after converting NOON state into single quantum coherence
(lower trace).
}
\label{tmp}
\end{center}
\end{figure}

Consider a star-topology spin system as shown in Fig. \ref{torrey}(b). 
Essentially it is an $AM_{N-1}$ spin system, with a single heteronucleus $A$ interacting with $N-1$ 
magnetically equivalent $M$ spins via J-coupling.  
An example system is shown in Fig. \ref{tmp}.
The spectrum of $A$ spin is split into
$N$ lines, depending on the spin-states of $M_{N-1}$ spins.  One of the extreme lines of this 
spectrum corresponds to the coherence 
$\frac{1}{\sqrt{2}}(\vert 0 \rangle + \vert 1 \rangle)\vert 00 \cdots 0 \rangle$, 
where each $M$ spin is in the state $\vert 0 \rangle$ (see Fig.2).  
This coherence can be converted into a NOON state \
using ${CNOT}$ gates ($NOT$ on \underline{all} $M$ spins controlled by $A$ spin),
\begin{eqnarray}
\frac{1}{\sqrt{2}}(\vert 0 \rangle + \vert 1 \rangle) \vert 00 \cdots 0\rangle 
&\stackrel{CNOT}{\longrightarrow}&
\frac{1}{\sqrt{2}}(\vert 00 \cdots 0 \rangle + \vert 11 \cdots 1 \rangle)
= \vert N00N \rangle.
\end{eqnarray}
Since NMR detection is confined to single quantum coherences, it is necessary
to convert the NOON state back to single quantum coherence before detection.
This can be achieved by the same set of $CNOT$ gates
\begin{eqnarray}
\vert N00N \rangle \stackrel{CNOT}{\longrightarrow}
\frac{1}{\sqrt{2}}(\vert 0 \rangle + \vert 1 \rangle) \vert 00 \cdots 0\rangle.
\end{eqnarray}
The pulse diagram for NOON state preparation and selection is shown in Fig. \ref{torrey}(c).
It can be observed that the $CNOT$ gates in star-topology systems can be implemeted
in parallel, and therefore require just two RF pulses on $M$ spins.  
A pair of PFGs, G$_1$ and G$_2$, serve to select the above coherence pathway, and extract 
the corresponding extreme spectral line (see Fig.\ref{tmp}).  The PFG ratio 
\begin{eqnarray}
\frac{G_2}{G_1} = (N-1)\frac{\gamma_M}{\gamma_A}+1,
\end{eqnarray}
depends on the size of the spin system ($N$) and the relative gyromagnetic ratio
$\gamma_M / \gamma_A$.

\subsection{Measurement of RFI via NOON states}
The pulse diagram for the measurement of RFI via NOON states is shown in Fig. \ref{torrey}(c).
In order to measure RFI without changing the coherence order of NOON states, we utilize
z-pulses.  
A $\phi_z$ pulse acting on a single spin-1/2 system introduces a relative phase shift
\begin{eqnarray}
\frac{1}{\sqrt{2}}(\vert 0 \rangle + \vert 1 \rangle)  \stackrel{\phi_z}{\longrightarrow} 
\frac{1}{\sqrt{2}}(\vert 0 \rangle + e^{i\phi_z} \vert 1 \rangle).
\end{eqnarray}
If the same pulse is applied on the $M_{N-1}$ spins of a NOON state, and subsequently converted into single
quantum coherence, we obtain
\begin{eqnarray}
\frac{1}{\sqrt{2}}(\vert 00 \cdots 0 \rangle + \vert 11 \cdots 1 \rangle)  
&\stackrel{\phi_z^M}{\longrightarrow}&
\frac{1}{\sqrt{2}}(\vert 00 \cdots 0 \rangle + e^{(N-1)i\phi_z} \vert 11 \cdots 1 \rangle)
\nonumber \\
&\stackrel{CNOT}{\longrightarrow} &
\frac{1}{\sqrt{2}}(\vert 0 \rangle + e^{i(N-1)\phi_z} \vert 1 \rangle) \vert 00 \cdots 0\rangle.
\end{eqnarray}
Therefore the resultant phase-shift of the extreme spectral line is 
$\phi_z^M = (N-1)\phi_z$.
A $\phi_z^M$ pulse, on $M$-spins can be realized by $\left(\frac{\pi}{2}\right)^M_{-y} \phi_x^M \left(\frac{\pi}{2}\right)_{y}^M$. Here
$\phi_x^M = 2\pi (N-1) \nu_0 \tau$ is the corresponding x-pulse of duration $\tau$
for a nominal RF amplitude $\nu_0$.
 Ideally,
in the absence of RFI, one should see a regular oscillatory behavior of this coherence with $\tau$.
In practice we see Torrey oscillation, i.e., decaying oscillations, due to the RFI during $\phi_x^M$
pulse.  As described in section 2.1, NOON Torrey-oscillations 
in large spin-systems decay much faster than the single-quantum Torrey oscillations.
The faster decay of NOON Torrey-oscillations allows one to study RFI at higher amplitudes.  

Here RFI of other short and fixed-duration pulses only introduce an over-all scaling factor, and 
do not contribute to the decay constant of signal.  Also, we assumed that the transverse relaxation
time constant of NOON state is much longer than the decay constant of NOON Torrey oscillations.

\subsection{Correlation between RFI of two channels}
In a two-channel probe, the regions of high RF intensity of first channel may
not correspond to regions of high RF intensity of the second.  In other words,
there exists certain correlations between the RFI profiles of the two channels.
The NOON state method can be easily
extended to study such correlations in the inhomogeneities of two different RF channels.
We choose $^1$H and $^{31}$P channels of Bruker QXI probe for studying such 
correlations and use trimethylphosphite (see Fig. \ref{tmp})
as the star-topology spin system ($^{31}P^{1}H_9$).  

\begin{figure}[b]
\begin{center}
\includegraphics[width=9.5cm]{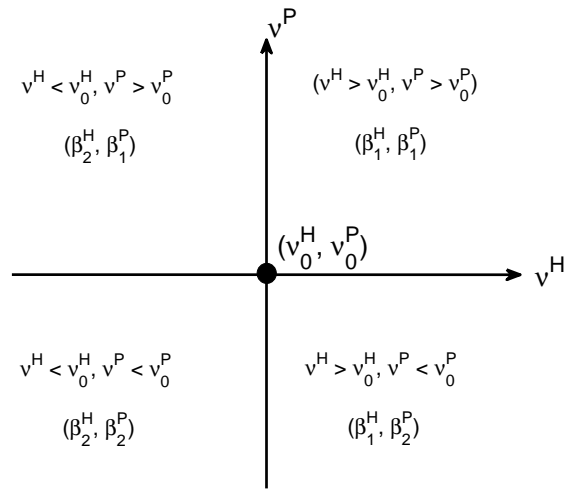}
\caption{Convention used for selection of $\beta$ parameters of the 3D asymmetric Lorentzian.
The central point $(\nu_0^H,\nu_0^P)$ indicates the nominal point, and the four quadrants
have different pairs of $\beta$ parameters.
}
\label{lambda}
\end{center}
\end{figure}

The pulse-diagram for the RFI-correlation study is shown in Fig. \ref{torrey}(d).
A $\phi_z^{P}$ pulse is introduced after the $\phi_z^{H}$ pulse, and the two pulses are independently
incremented to obtain a 3D dataset.  
Similar to Eq. \ref{snu}, the Fourier transform along the two dimensions result
in signal $S(\nu^H_9,\nu^P_1)$.  We now model the RFI correlation using a 3D asymmetric
Lorentzian:
\begin{eqnarray}
p(\nu^H,\nu^P) = 
\frac{\lambda_0^2}{d^2(\nu^H,\nu^P) + \lambda_0^2},
\end{eqnarray}
where $d(\nu^H,\nu^P)$ is the scaled distance of 
$\left(\frac{\nu^H}{\nu_0^H},\frac{\nu^P}{\nu_0^P}\right)$ from the nominal point 
$(1,1)$:
\begin{eqnarray}
d^2(\nu^H,\nu^P) = \beta^H_i \left( 1 - \frac{\nu^H}{\nu^H_0} \right)^2 
        + \beta^P_j \left(1 - \frac{\nu^P}{\nu^P_0} \right)^2.
\end{eqnarray}
Here $\beta^H_1$, $\beta^H_2$, $\beta^P_1$, and $\beta^P_2$ are the
four asymmetry parameters, on the four quadrants of $\nu^H$ - $\nu^P$
plane (Fig. \ref{lambda}).  The probability distribution is normalized to unit sum.

As described in section 2.1, we first constrcut the model torrey oscillation 
$s_m(\tau^H,\tau^P)$ and its Fourier transform $S_m(\nu^H_9,\nu^P_1)$.  By fitting
$S_m(\nu^H_9,\nu^P_1)$ with the experimental profile $S(\nu^H_9,\nu^P_1)$, we obtain
the best fit parameters $(\lambda_0,\{\beta\})$.

In the following, we describe experimental characterizations of 
(i) $^1$H RFI at different RF powers in two NMR probes, and 
(ii) RFI correlation between $^1$H and $^{31}$P channels of a NMR probe.

\section{Experiments}
The single-quantum Torrey oscillations were studied using a sample consisting of 
600 $\upmu$l of 99\% D$_2$O.  The NOON Torrey oscillations were studied using 
100 $\upmu$l of trimethylphosphite (Fig. \ref{tmp}) dissolved in 500 $\upmu$l of dimethylsulphoxide-D6.  
The J-couplings between the magnetically equivalent methyl protons and the $^{31}$P spin are all equal 
to 11 Hz. The effective T$_2^*$ relaxation time constants of $^1$H and $^{31}$P are about 150 ms 
and 730 ms respectively (as is the general case, T$_{1\rho}$ is longer than T$_2^*$).  These values
were obtained from respective spectral linewidths.  The NOON state T$_2^*$ is about 38 ms, and is
obtained by introducing a variable delay between the two CNOTs, 
and then measuring the decay constant of the single quantum coherence after conversion.
All the experiments were carried 
out on a Bruker 500 MHz NMR spectrometer at an ambient temperature of 300 K. 

\begin{figure}
\begin{center}
\hspace*{-1.5cm}
\includegraphics[width=13cm]{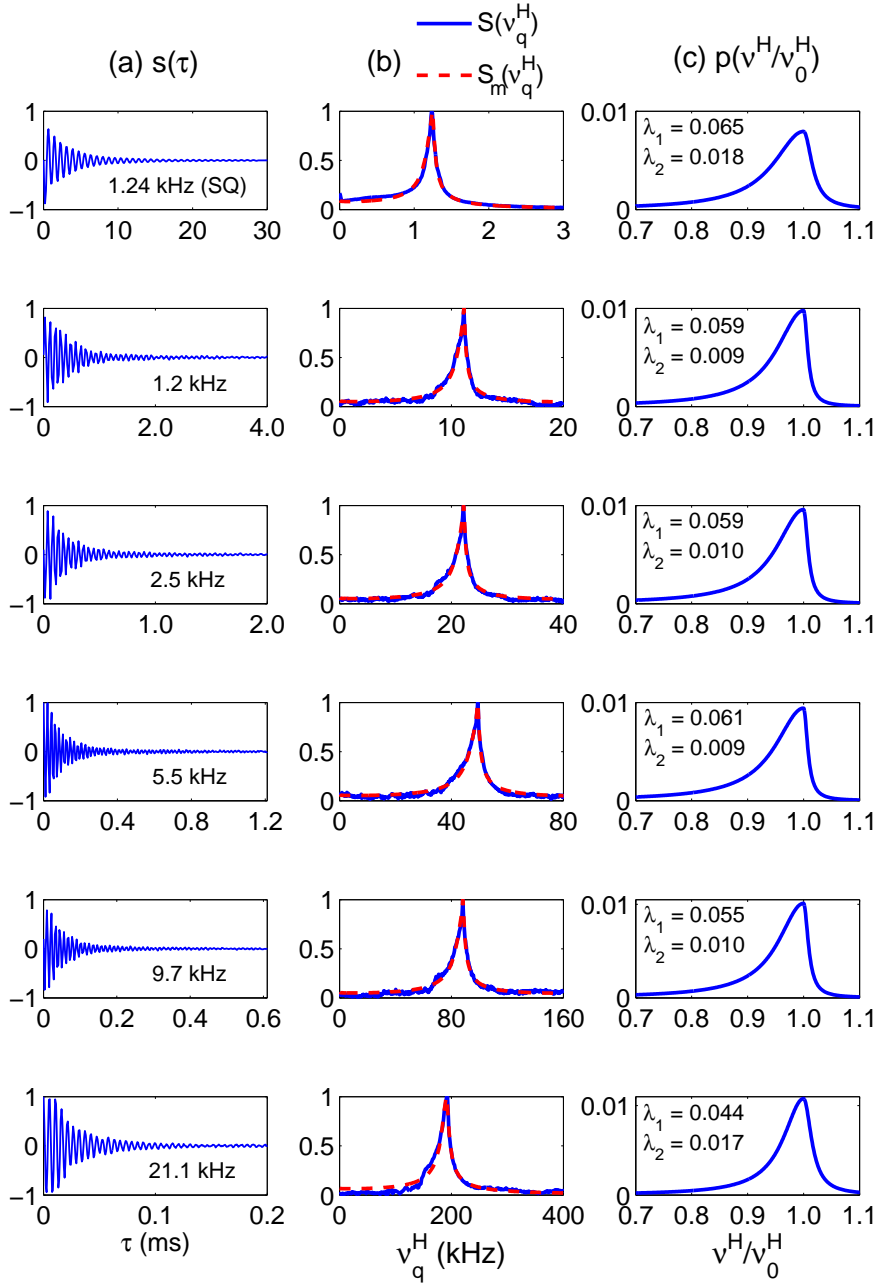}
\caption{ Characterization of RFI in $^1$H channel of BBO probe.
The results with single quantum method are in the top trace, and other traces are for
NOON states with different nominal RF amplitudes (as indicated in column (a)).
Column (a): Torrey oscillations s($\tau$) vs $\tau$.
Column (b): Positive real part of Fourier transform $S(\nu^H_q)$ vs $\nu_q^H$ (solid lines).  
Also shown are the best fits with model $S_m(\nu^H_q)$ (dashed lines).
Column (c): The RFI profiles p($\nu^H/\nu_{0}^H$) vs $\nu^H/\nu_{0}^H$. 
The asymmetric linewidth parameters $(\lambda_1,\lambda_2)$ are indicated in each case.
}
\label{bboplotall}
\end{center}
\end{figure}

\begin{figure}
\begin{center}
\hspace*{-1.5cm}
\includegraphics[width=13cm]{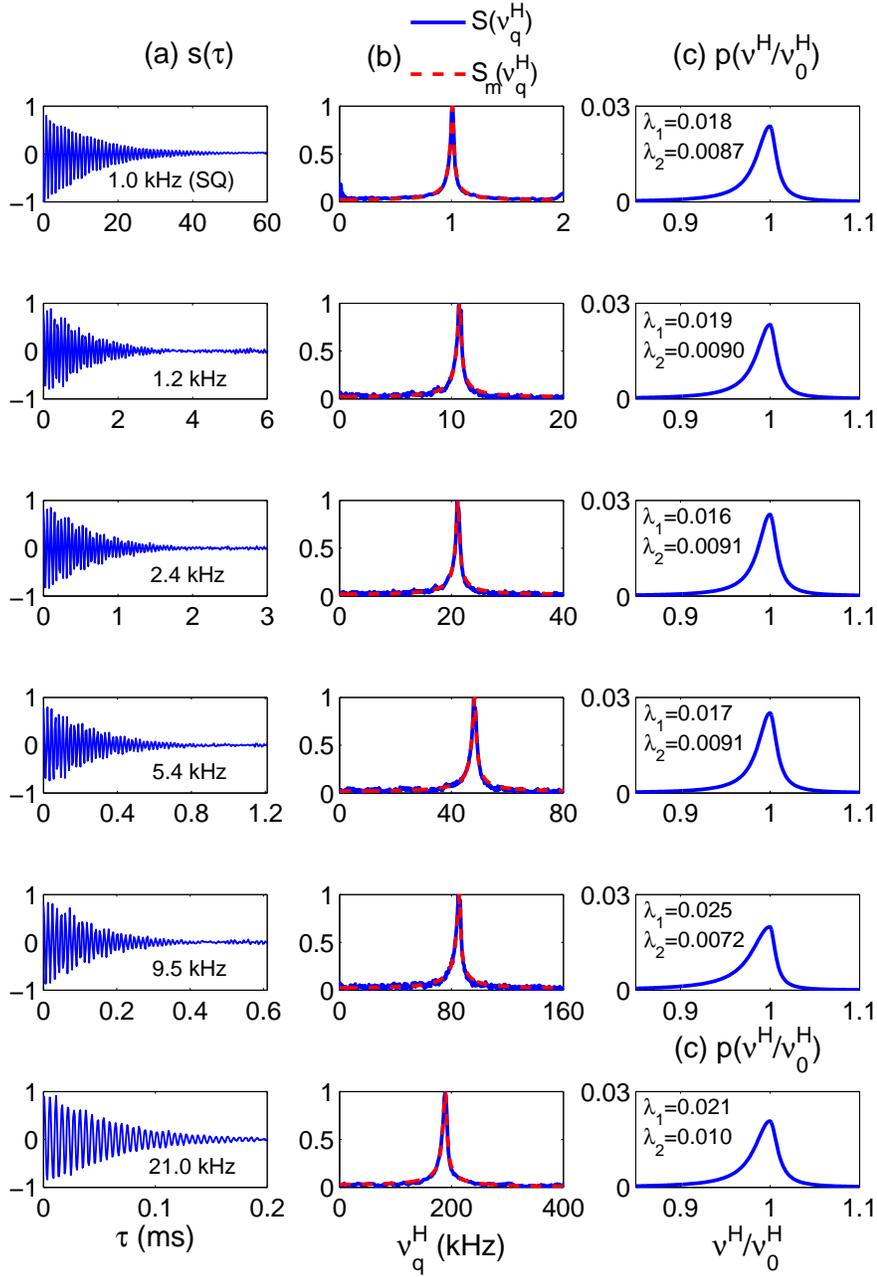}
\caption{ Characterization of RFI in $^1$H channel of QXI probe.
The results with single quantum method are in the top trace, and other traces are for
NOON states with different nominal RF amplitudes (as indicated in column (a)).
Column (a): Torrey oscillations s($\tau$) vs $\tau$.
Column (b): Positive real part of Fourier transform $S(\nu^H_q)$ vs $\nu_q^H$ (solid lines).  
Also shown are the best fits with model $S_m(\nu^H_q)$ (dashed lines).
Column (c): The RFI profiles p($\nu^H/\nu_{0}^H$) vs $\nu^H/\nu_{0}^H$. 
The asymmetric linewidth parameters $(\lambda_1,\lambda_2)$ are indicated in each case.
}
\label{qxiplotall}
\end{center}
\end{figure}

\subsection{RFI characterization of $^1$H channel}
The single quantum Torrey oscillations were studied using the pulse sequence shown
in Fig. 1(a).  A series of spectral lines were recorded by incrementing the duration
of the low-power (about 1 kHz) on-resonant pulse on 99\% D$_2$O sample as described in section 2.1.
The results of the single quantum experiments on $^1$H channels of BBO and QXI probes 
are shown in the top trace of Figs. \ref{bboplotall} and \ref{qxiplotall}
respectively.  The nominal RF frequencies were found to be slightly different for
the two probes as indicated in the figures.  The $\tau$-increments in BBO and QXI were
set to 112.5 $\upmu$s and 250 $\upmu$s respectively, and a total of 256 transients were
recorded.  The Torrey oscillation ($s(\tau)$), the positive real part of the Fourier 
transform ($S(\nu_1^H)$) are shown in the first traces of Figs. \ref{bboplotall}(a,b) and 
\ref{qxiplotall}(a,b). By fitting the model $S_m(\nu_1^H)$ to $S(\nu_1^H)$,
we obtain the asymmetry parameters $\lambda_1$ and $\lambda_2$ and the corresponding
RFI profile $p(\nu^H)$.  The top traces of Figs. \ref{bboplotall}(c) 
and \ref{qxiplotall}(c) show the RFI profiles $p(\nu^H)$ and the corresponding $\lambda$ values for 
BBO and QXI probes respectively.

The NOON state is prepared with the $^{31}$P-$^1$H$_9$ system of trimethylphosphite (Fig. \ref{tmp})
using the pulse sequence shown in Fig. \ref{torrey}(c).  The results of the NOON Torrey oscillations
at different nominal amplitudes of $^1$H channels of BBO and QXI probes are shown in Figs. \ref{bboplotall}(a)
and \ref{qxiplotall}(a).  It can be observed that as the nominal amplitude is increased, the Torrey
oscillations decay faster.  The positive real parts of the Fourier transform $S(\nu_9^H)$ of the 
9-quantum Torrey oscillations are shown in Figs. \ref{bboplotall}(b) and \ref{qxiplotall}(b).  By 
fitting the model function $S_m(\nu^H_9)$ to $S(\nu_9^H)$ at each of these nominal
amplitudes, we obtained the corresponding RFI profiles $p(\nu^H)$.  These profiles and the 
asymmetry parameters $(\lambda_1,\lambda_2)$ are shown in Figs. \ref{bboplotall}(c)
and \ref{qxiplotall}(c).

It can be observed that decay times of the NOON state Torrey oscillations are
an order of magnitude shorter than the single quantum Torrey oscillations
in both probes.  This fact allowed us to characterize RFI safely even
at maximum allowed RF powers.
The excellent fit in all the cases indicates the validity of our model.
In the overall comparison, we find that the BBO probe has stronger RFI than QXI probe
at all RF amplitudes.  The mean values of $(\lambda_1,\lambda_2)$ are (0.057,0.012)
and (0.019,0.009) respectively in BBO and QXI probes.

%In BBO probe, we find that the RFI profiles obtained via NOON state at different 
%nominal RF amplitudes were similar, but slightly sharper compared to that obtained
%by the standard single-quantum method.  We believe that this slightly discrepancy arises due to the
%selection gradients employed in the NOON state method. The selection gradients effectively
%reduce the volume of the sample contributing to the final signal.
In QXI probe, the RFI profiles are not only sharper than those in BBO, but are also 
more symmetric.  The mean ratios $\lambda_1/\lambda_2$ are 5.2 and 2.2 in BBO and
QXI respectively.
QXI probe shows some what stronger RFI at higher RF amplitudes.  For example,
RFI profiles at 9.5 kHz and 21 kHz are significantly wider than the 
profile at 5.4 kHz.

\subsection{RFI correlation between $^1$H and $^{31}$P channels}
\begin{figure}
\begin{center}
\hspace*{-1cm}
\includegraphics[width=12.5cm,angle=-90]{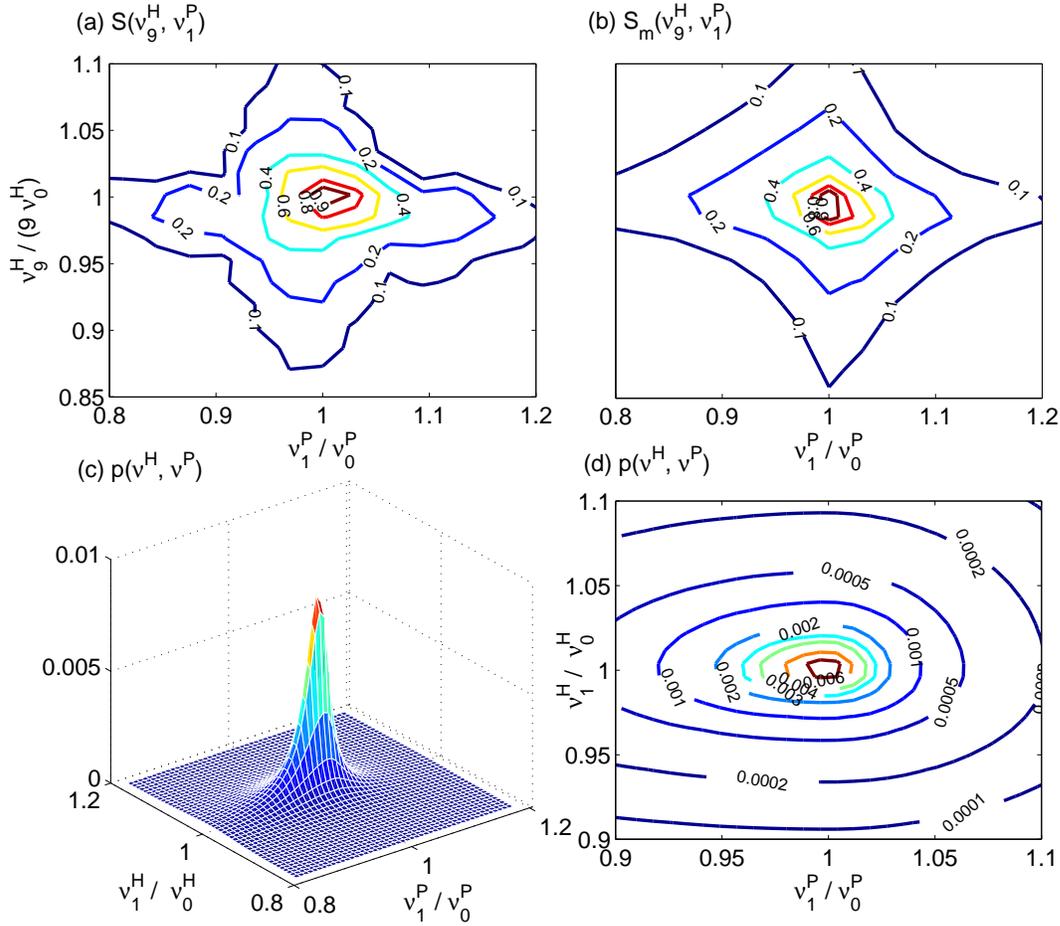}
\caption{Characterization of RFI correlations between $^1$H and $^{31}$P channels of
QXI probe. (a) Contour plot showing real positive part of Fourier transform of 2D 
Torrey oscillations $S(\nu^H_9,\nu_1^P)$, (b) corresponding simulated profile obtained
with the best fit model $S_m(\nu^H_9,\nu_1^P)$, (c) the surface plot of the RFI
profile $p(\nu^H,\nu^P)$, and (d) the contour plot of $p(\nu^H,\nu^P)$.
}
\label{rficorr}
\end{center}
\end{figure}

We measured RFI correlation between $^1$H and $^{31}$P channels of Bruker QXI probe
using the NOON state method.
The pulse sequence for measuring RFI correlation between $^1$H and $^{31}$P 
channels is shown in Fig. \ref{torrey}(d) and the theory is described in section 2.4.
The dataset of 2D Torrey oscillations was recorded by independently incrementing
the $\tau^H$ and $\tau^P$ pulses respectively by 11.1 $\upmu$s and 50.0 $\upmu$s.
The nominal RF amplitudes in $^1$H and $^{31}$P channels were 2.4 kHz and 2.5 kHz 
respectively.
A total of 128 data points in $^1$H dimension and 96 data points in $^{31}$P dimension were
collected.  The real positive part of the Fourier transform of the 2D Torrey oscillations 
$S(\nu^H_9,\nu^P_1)$ was obtained after a zero-fill to 256 points in each dimension.
The contour plot of $S(\nu^H_9,\nu^P_1)$ is shown in Fig. \ref{rficorr}(a). The asymmetric
Lorentzian parameters $(\lambda_0,\{\beta\})$ were obtained by fitting the model profile
$S_m(\nu^H_9,\nu^P_1)$ to $S(\nu^H_9,\nu^P_1)$.  Fig. \ref{rficorr}(b) displays 
$S_m(\nu^H_9,\nu^P_1)$ corresponding to the best fit:
$\{\lambda_0 = 0.0045, \beta_1^H = 0.114, \beta_2^H = 0.226, \beta_1^P = 0.095, \beta_2^P = 0.028\}$.
It can be seen that the model profile $S_m(\nu^H_9,\nu^P_1)$ approximates the overall profile of $S(\nu^H_9,\nu^P_1)$.
The corresponding RFI profile $p(\nu^H,\nu^P)$ is shown in Fig. \ref{rficorr}(c).
It can be observed that $^{31}$P-channel has a wider RFI distribution than the $^1$H-channel.

\section{Conclusions}
We have proposed an asymmetric Lorentzian model to describe radio-frequency
inhomogeneity.  Using this model, we obtained the asymmetric line-width 
parameters in two high-resolution NMR probes using single quantum Torrey
oscillations.  The excellent fit between the experimental profile and the simulated
profile indicated the suitability of the asymmetric Lorentzian model.

We have described the characterization of RFI using NOON states.  
NOON state Torrey oscillations decay much faster than single quantum Torrey
oscillations allowing the characterization of RFI at higher RF amplitudes.
Using this method, we have studied RFI of two NMR probes at different RF amplitudes and 
compared the results.  
We have also proposed a 3D Lorentzian model to describe RFI correlations 
between two RF channls. Then we extended the NOON state method 
to characterize such correlations, and demonstrated it on $^1$H-$^{31}$P channels.
We believe that these characterizations will be useful to design
high-precision NMR and MRI experiments as well as to design high-fidelity
quantum gates required in NMR quantum information processing.

\section*{Acknowledgments}
The authors are grateful to Prof. Anil Kumar and Soumya Singha Roy for
discussions. This work was partly supported by the DST project SR/S2/LOP-0017/2009.

\end{document}